\documentclass[aps,reprint]{revtex4-1}

\usepackage{graphicx}

\usepackage{mathrsfs}

\usepackage{hyperref}
\usepackage[dvipsnames]{xcolor}
\newcommand\myshade{95}
\colorlet{mylinkcolor}{BrickRed}
\colorlet{mycitecolor}{NavyBlue}
\colorlet{myurlcolor}{BrickRed}
\hypersetup{
  linkcolor  = mylinkcolor!\myshade!black,
  citecolor  = mycitecolor!\myshade!black,
  urlcolor   = myurlcolor!\myshade!black,
  colorlinks = true,
}

\setcitestyle{authoryear,open={(},close={)}} 

\begin{document}

\title{Random State Comonads encode Cellular Automata evaluation}

\author{Madalina I Sas}
\affiliation{Centre for Complexity Science, Imperial College London, UK}

\author{Julian HJ Sutherland}
\affiliation{Nethermind, UK}

\begin{abstract}
    \noindent Cellular automata (CA) are quintessential ALife and ubiquitous in many studies of collective behaviour and emergence, from morphogenesis to social dynamics and even brain modelling. 
    Recently, there has been an increased interest in formalising CA, theoretically through category theory and practically in terms of a functional programming paradigm. 
    Unfortunately, these remain either in the realm of simple implementations lacking important practical features, 
    or too abstract and conceptually inaccessible to be useful to the ALife community at large.
    In this paper, we present a brief and accessible introduction to a category-theoretical model of CA computation through a practical implementation in Haskell.
    We instantiate arrays as comonads with state and random generators, allowing stochastic behaviour not currently supported in other known implementations.
    We also emphasise the importance of functional implementations for complex systems: thanks to the Curry-Howard-Lambek isomorphism, functional programs facilitate a mapping between simulation, system rules or semantics, and categorical descriptions, which may advance our understanding and development of generalised theories of emergent behaviour. 
    Using this implementation, we show case studies of four famous CA models: first Wolfram's CA in 1D, then Conway’s game of life, Greenberg-Hasings excitable cells, and the stochastic Forest Fire model in 2D, 
    and present directions for an extension to $N$ dimensions.
    Finally, we suggest that the comonadic model can encode arbitrary topologies and propose future directions for a comonadic network.
\end{abstract}

\maketitle

\section{Introduction}

\noindent The origins of life remain mysterious. As mysterious is the ability of artificial life to imitate emergent life-like properties and behaviours, and to do so by computation. 
As such, ALife is intrinsically linked to computer science, as in the study of computation, and to complexity science, as in the study of emergence.

All life can be seen as a complex system, 
made up of a multitude of components which interact in simple but often non-linear ways,
showing emergent properties or behaviours as the result of self-organisation \citep{jensen_2022}.
Emergence in complex systems is often studied by modelling these interactive processes and creating the simplest simulation that can reproduce the emergent behaviour at the system level. 
Phenomena reproducible in this way are often deemed \textit{weakly} emergent \citep{bedau_2008}:
this method bypasses the explanatory gap of strong emergentist accounts, but also limits the phenomena explored to those that can be expressed as \textit{computation}.

\begin{figure}
    \centering
    \includegraphics[width=\linewidth]{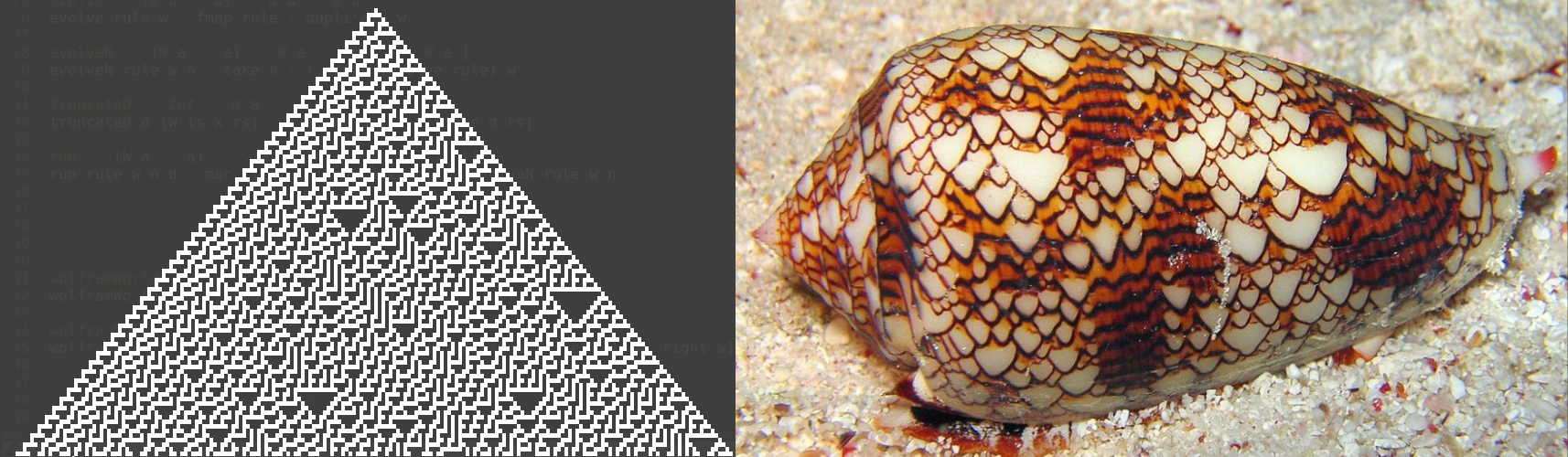}
    \caption{\textbf{Artificial versus natural emergent patterns.}\\
 (a) A Wolfram CA using rule 30 and an initial world with only one live cell in the middle. (b) The rule 30 CA shows very similar patterns to the \textit{Conus textile} shell.
    }
    \label{fig:wolfram-conus-tex}
    \vspace{-1.5em}
\end{figure}

Cellular automata (CA) are an obvious example of such computational emergent behaviour, and can model a broad range of real-world phenomena.
Being discrete in time and space, they allow the utter simplification of the interactive processes between parts. Emergence manifests even in a 1D model such as the one proposed by \citet{wolfram_1983}, where a trivial rule can produce pure unpredictable chaos, with such computational complexity it has puzzled cryptanalysts~\citep{wolfram_2019},
while maintaining spatial patterns uncannily similar to those seen in nature (see Fig \ref{fig:wolfram-conus-tex}).

Given this simplicity, it is tempting to try to express the rules underlying the interactions between cells into a mathematical theory that describes their behaviour, 
but also into a practical implementation that is intrinsically linked to theory.
This idea is, of course, not new: 
there is significant historical and recent work establishing links between category theory and automata theory,
by using categorical structures to model, operationalise, and generalize automata and learning algorithms
\citep{rine_1971,vanheerdt2020calf,colcombet2020learning}.
These methods allow us to identify the theoretic capabilities of a general interactive system based on descriptions of its input and output states, but depend on finding the right description which is often non-trivial.
In the case of CA, a number of novel algebraic or categorical descriptions have been recently developed \citep{capobianco_uustalu_2010, widemann_hauhs_2011,widemann_2012,basold_2025}.

Implementation-wise, the functional programming paradigm appears particularly well-suited: 
its emphasis on immutability and pure functions lacking side-effects maps to the deterministic and iterative aspects of dynamical systems, facilitates reproducibility, but also naturally aids parallelisation.
Moreover, strongly-typed functional languages, by design, enforce mathematical constraints at compilation, guaranteeing correct execution, and lend themselves towards direct correspondences to semantics.
As such, the functional programming community has been taking an interest in this problem for more than a decade \citep{piponi_2006,potts_2006,
kmett_2015,bollu_2019_gh}, but in the scientific modelling of multi-agent systems most implementations remain procedural and array-based.

However, existing algebraic descriptions for CA are often too abstract to be yet of use to the ALife community at large;
meanwhile, relevant practical aspects, such as the stochasticity required by the Forest Fire model,
remain unadressed in existing functional implementations.

The functional programming approach to CA could also expand our general understanding of emergent phenomena, by formalising a mapping between semantics, mathematical descriptions and simulations of complex systems.
With this motivation, this work aims to reunite CA, category theory and a functional approach through an implementation in Haskell using random-comonadic arrays, and presents some case studies including stochastic CA.

\section{Preliminaries}\label{sec:prelim}

\noindent At its core, a CA can be seen as executing repeated, context-dependent computations: each cell evolves based on its position in the grid, and the states of itself and its neighbours. 
Category theory provides abstract mathematical structures that can encode this evolution regardless of model specifics.
In this section, we present the background to expressing CA both as categories and as functional programs. 

\subsection{Functional programming and type theory}

\noindent Functional and procedural programming offer distinct paradigms in their approach to computation,
following the two distinct conceptual frameworks supporting the Church-Turing thesis.
Turing machines treat computation as a sequence of transitions between states expressed as symbol manipulations on a tape, emphasising \textit{how} to compute \citep{turing1936} 
-- naturally expressing step-by-step procedural programming.
Albeit initially less intuitive or practical, Church’s $\lambda$-calculus focuses on \textit{what} to compute, representing computation as function application and composition \citep{church1936} -- inspiring declarative, function-centered programming. 

In typed $\lambda$-calculus, each expression is further enriched by an associated \textit{type}.
The correctness of programs is ensured by enforcing strict rules on how types can be used.
While this reduces the expressive power of $\lambda$-calculus, 
programs in typed $\lambda$-calculus can be mapped directly to mathematical structures \citep{Mitchell1990}.

Furthermore, \textit{the Curry-Howard isomorphism} allows mapping types to logical propositions, effectively providing a correspondence between type-checking and program verification \citep{hoare1980}, which allows for extra safety guarantees when compiling strongly-typed languages.

\subsection{From types to categories}

\noindent  \textit{"Category theory takes a bird's eye view of mathematics"} \citep{leinster_2017}, allowing very general descriptions of abstract mathematical structures and the relationships between them. 
Category theory can provide a unifying paradigm to model automata, enabling generic definitions and descriptions of their behaviour \citep{rine_1971}.

A \textit{category} $\mathscr{C}$ consists of a collection of objects and relationships (\textit{morphisms} or \textit{arrows}) between them. Each arrow has a domain and codomain, extracted from the objects in $\mathscr C$. Morphisms can be chained using an associative composition operator $\circ$. An identity morphism maps each object $A$ in $\mathscr C$ to itself Id$_A: A \rightarrow A$.

Category theory provides a framework to express the semantics of typed $\lambda$-calculus: by modelling types as objects in a category, and functions as morphisms between objects.
Then, compositions of morphisms corresponds to function composition, and identity morphisms correspond to identity functions.
Through the above, the correspondence between programs and logic can also be expanded to categories; this is known as the \textit{Curry-Howard-Lambek isomorphism}.

A \textit{functor} $F: \mathscr C\rightarrow \mathscr D$ is a higher-order concept: a mapping that preserves structure 
between two categories $\mathscr C$ and $\mathscr D$, by assigning to each object and each morphism in $\mathscr C$ an object and morphism in $\mathscr D$ in a way that maintains composition and identity of morphisms.
In particular, a functor $E: \mathscr C \rightarrow \mathscr C$ can map a category to itself; such \textit{endofunctors} can be equipped with constraints and properties which can be used to intuitively encode certain kinds of computation.

A \textit{monad} is an endofunctor $M:\mathscr C\rightarrow \mathscr C$ with two natural transformations: a unit element $e: I \to M$ and a multiplication $\mu: M \times M \to M$, satisfying identity and associativity laws analogous to those of a monoid.

A \textit{comonad} is defined by `reversing the arrows' in the monad: it is an endofunctor with a \textit{co}unit $\varepsilon: W \to I$ and \textit{co}multiplication $\delta: W \to W \times W$. 

Intuitively, a monad maps two elements to one, while a comonad maps an element to two. Monads and comonads have a \textit{dual} relationship, such that the types of their fundamental operators are reversed. When thinking of computation, the former can act as an analogy for chaining functions sequentially, while the latter can encapsulate data extraction from a larger context \citep{uustalu_2008}.

\subsection{Implementation in Haskell}

\noindent Haskell is a strongly-typed, functional programming language which implements many concepts in both $\lambda$-calculus and category theory.

Loosely, types correspond to objects in a category, while functions of type \textsf{a} $\to$ \textsf{b} correspond to morphisms between those objects of types \textsf{a} and \textsf{b}.

Endofunctors can be written using a data constructor, such as \textsf{f a}, for an endofunctor \textsf{f} and an arbitrary type \textsf{a}. 
Practically, functors can be seen as data structures that allow mapping a function (or morphism) to their elements. Lists, for example, are functors: \textsf{map} takes a function and a list and applies it to all elements, returning an updated list: 

\vspace{0.3em}
\textsf{map\; :: (a $\to$ b) $\to$ [a] $\to$ [b]}

\textsf{fmap :: Functor f $\Rightarrow$ (a $\to$ b) $\to$ f a $\to$ f b}
\vspace{0.3em}

\noindent The comonad can also express a kind of container or data structure for some arbitrary type.
The comonadic operators $\epsilon: W \to I$ and $\delta: W \to W \times W$ map to the functions \textsf{extract} and \textsf{duplicate}:

\vspace{0.3em}
\textsf{extract \quad :: Comonad w $\Rightarrow$ w c $\to$ c}

\textsf{duplicate \. :: Comonad w $\Rightarrow$ w c $\to$ w (w c)}
\vspace{0.3em}

\section{A comonadic model of Cellular Automata}\label{sec:model}

\noindent  Let us now consider a cell of type \textsf{c} and a comonad \textsf{w} that represents a grid of cells.
We compute the value of each cell in the next generation by performing a\textit{ local} computation based on its neighbours. A comonad allows us to \textit{focus} on each individual cell in the grid, while the value of its neighbours and the rest of the grid are considered part of the \textit{context}.
A full category-theoretical definition of comonadic CA is available in \citet{capobianco_uustalu_2010}.
For simplicity, we will present the model using function types and data constructors in Haskell.

Most comonadic implementations of CA enhance the list functor with a context in the form of a list zipper data structure, which has a focused cell and two lists left and right for the remaining cells: \textsf{W [c] c [c]}. Checking the value of neighbours is done by context-switching: navigating the focus point to a neighbour -- which can also make implementing periodic boundaries trivial. 

However, lists are inefficient to index, making the expansion to multiple dimensions difficult. Taking inspiration from procedural implementations, we propose that an $n$-dimensional array, which is a kind of functor, features comonadic structure when considered with an index \textsf{i} to choose which element is currently in focus.

\vspace{0.3em}
\textsf{data W i c = W i (Array i c) deriving Functor}
\vspace{0.3em}

\noindent This can be shown by turning the array datatype into a comonad via an instance, which entails mapping the comonadic operators \textsf{extract} and \textsf{duplicate} to equivalent functions on arrays:
\textsf{extract} simply returns the value of the focused cell at index \textsf{i} from the array,
while \textsf{duplicate} takes an array focused on element at index \textsf{i}, and creates a an array of arrays expressing a `grid-of-grids' where each possible index is in focus.

\vspace{0.3em}
\textsf{instance Ix i $\Rightarrow$ Comonad (W i) where}

\textsf{\quad extract (W i a) = a ! i }

\textsf{\quad duplicate (W i a) = U i (listArray ix grids)}

\textsf{\quad \quad where}
        
\textsf{\quad\quad\quad ix = bounds a}

\textsf{\quad\quad\quad grids = [W i a $|$ i $\leftarrow$ range ix]}

\vspace{0.3em}

\noindent Suppose we then choose a rule by which a cell evolves according to its neighbours. The rule would take relevant neighbour states from the whole grid of type \textsf{w c}, but return only a cell of type \textsf{c}, with the same type as \textsf{extract}:

\vspace{0.3em}
\textsf{rule \quad :: Comonad w $\Rightarrow$ w c $\to$ c}
\vspace{0.3em}

\noindent In order to apply this rule everywhere in the grid, we need to apply it to each possible shift of the grid, for which we use \textsf{duplicate}. 
The \textsf{extend} function allows us to apply evolution rules \textsf{r} over the `duplicated' system state \textsf{s}, effectively 'reducing' the nested grid-of-grids structure of type \textsf{w (w c)} back to a grid of type \textsf{w c} one generation later:

\vspace{0.3em}
\textsf{extend :: (w c $\to$ c) $\to$ w c $\to$ w c}

\textsf{extend r s = fmap r \$ duplicate s}
\vspace{0.3em}

\noindent This elegant implementation is particularly general as there are no constraints on the structure of the index: the index type \textsf{Ix i} can be instantiated to either an integer, for 1D arrays, a pair for 2D arrays, or a larger tuple or a fixed-length list for arbitrary dimensions.

\subsection{Transform to random}

\noindent  Being able to add random computations is crucial to many multi-agent simulations. 
To implement this in a pure languages, the side-effects of stochasticity can be contained in a monad, allowing reproducibility based on random seeds. 
We also use the random monad to create random initial conditions, by making a cell an instance of the \textsf{Random} class and using a random number generator.

To allow stochasticity at each iteration, two transformations need be applied to the comonad datatype.
First, we use an environment comonad transformer, \textsf{EnvT}, which allows us to carry-on any model parameters as state, and the random monad transformer, \textsf{RandT}, which allows us to contain the side-effects of a random number generator with a seed \textsf{g}:

\vspace{0.3em}
\textsf{data RandGrid p i c =} 

\textsf{\quad RandT StdGen (EnvT Params (U i)) c}
\vspace{0.3em}

\noindent Using this data structure, we can apply various array dimensions and indices, cell datatypes, or simulation parameters, allowing for a very versatile range of simulations, as shown by the case studies to follow.

Our full implementation is available online \citep{sas_sutherland_2019}
and uses the \textsf{Comonad} package \citep{kmett_2-14}.

\section{Case Studies}\label{sec:case}
Below, we present four case studies of CA of increasing semantic complexity, with simulations produced by the implementation described above.  

\subsection{Wolfram CA}

\vspace{-0.5em}

The one-dimensional CA proposed by \citet{wolfram_1983} is usually the starting point of all functional implementations of CA (such as \citet{piponi_2006}).
Cell values are binary, and its evolution rules are given by a function applied to value of the current cell and its two immediate neighbours, with some combinations producing unexpectedly complex outcomes. In this example we use rule 22, which produces the famous Sierpinski triangle.

\begin{figure}[h!]
    \centering
    \includegraphics[width=\linewidth]{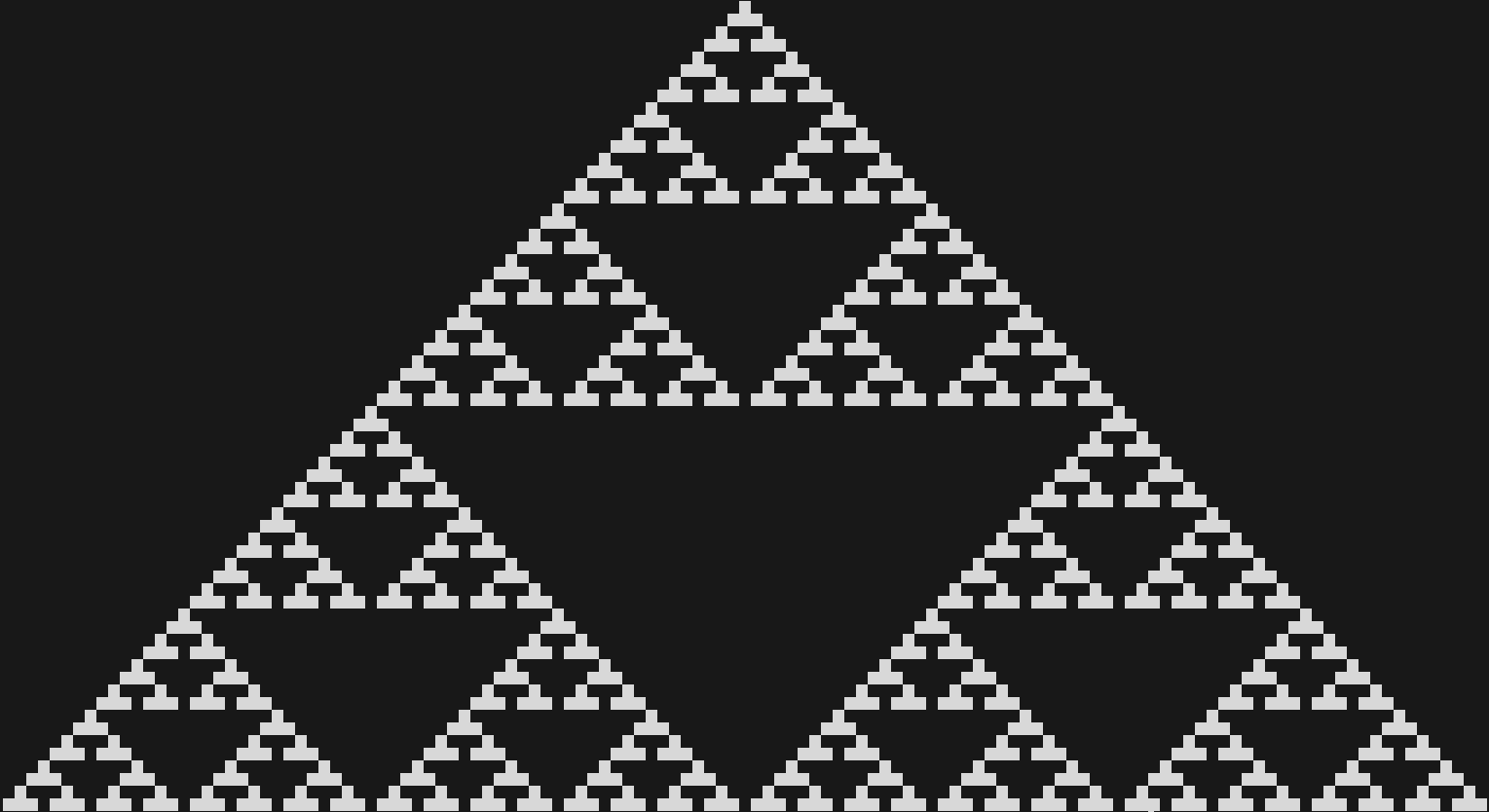}
    \caption{\textbf{Simulation of Wolfram rule 22}.
Consecutive snapshots from a 65 cell system, with initial conditions given by one living cell in the middle.
    }
    \label{fig:wolfram}
    \vspace{-1.5em}
\end{figure}

\subsection{Conway's Game of Life}

\vspace{-0.5em}

The Game of Life needs no introduction. We use the comonadic 2-dimensional array to encode encode the world, and a binary cell type. Local interactions are encoded using all 8 neighbours in 2D, with a cell surviving or being spawned if and only if it has two or three living neighbours. 

\begin{figure}[h!]
    \centering
    \includegraphics[width=0.24\linewidth]{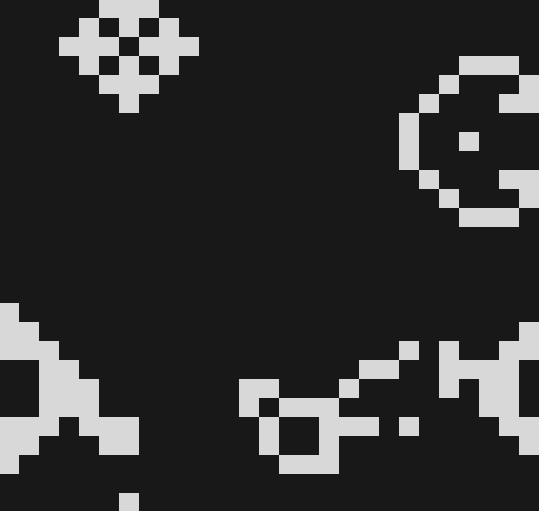} \includegraphics[width=0.24\linewidth]{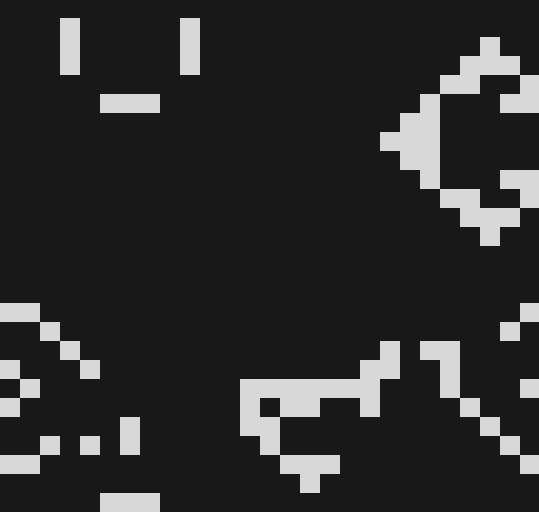} \includegraphics[width=0.24\linewidth]{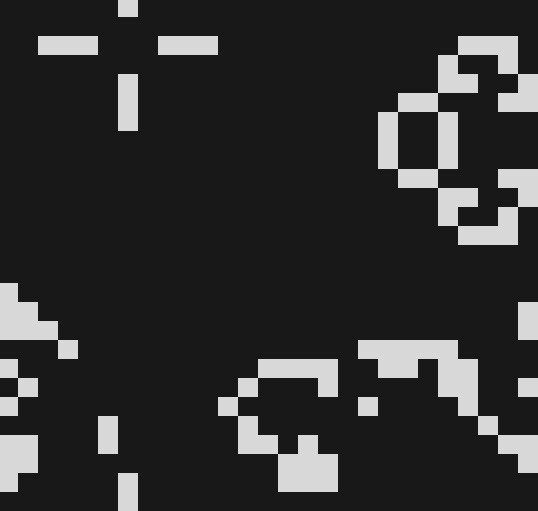} \includegraphics[width=0.24\linewidth]{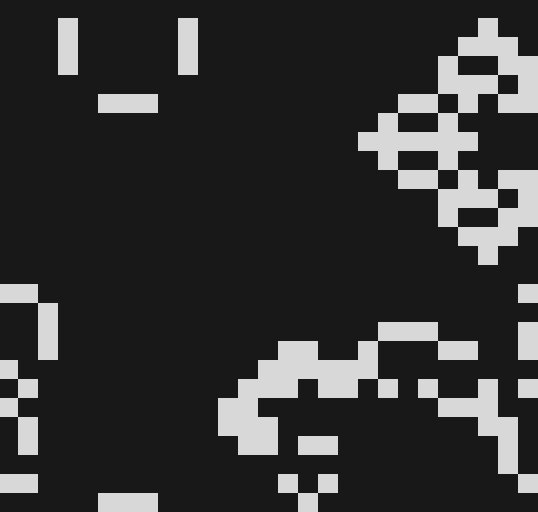}
    \caption{\textbf{Simulation of Game of Life }.
Consecutive snapshots from a 27x27 system with random initial conditions (seed $s_1=4123$) and $p=0.1$ of being alive. A `Traffic Light' period-2 pattern is recognisable in the top-left corner. }
    \label{fig:conway}
    \vspace{-1.5em}
\end{figure}

\subsection{Greenberg-Hasting excitable media}

\vspace{-0.5em}

Another classic model, the excitable medium, introduces a cell type inspired by neural spiking and slime moulds \citep{greenberg_hastings_1978}. The cell type has three possible states: \textsf{Quiet}, \textsf{Spike}, and \textsf{Rest}. Quiet cells will spike if there is a nearby spiking cell; after spiking, cells rest in a refractory state, during which they do not respond to stimulation; resting cells then return to the equilibrium quiet state.

In our simulation, we started with random initial configurations. If only spiking and quiet cells are included the grid quickly returns to the quiet equilibrium state.
But when including at least one resting cell, travelling waves of activity emerge, with the system self-organising into periodic behaviour, which reminds of models of epilepsy \citep{rabinovitch_2024} and atrial fibrillation \citep{ciacci_falkenberg_2020}.

\begin{figure}[h!]
    \centering
    \includegraphics[width=0.24\linewidth]{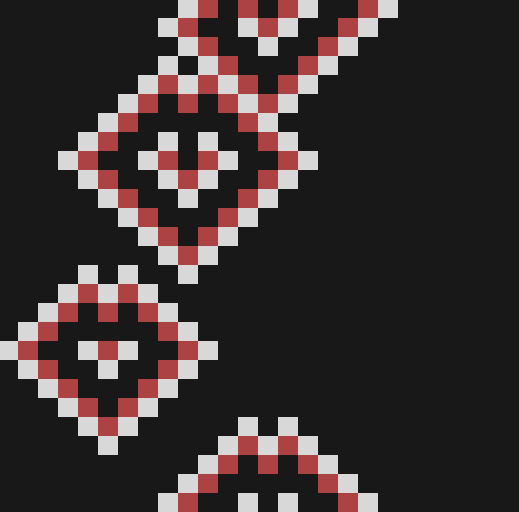} \includegraphics[width=0.24\linewidth]{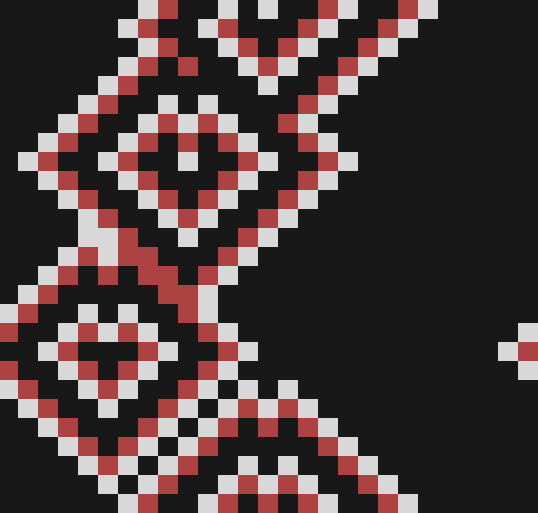} \includegraphics[width=0.24\linewidth]{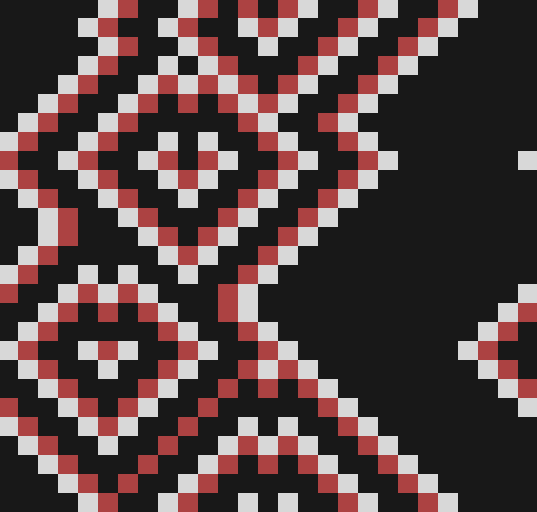} \includegraphics[width=0.24\linewidth]{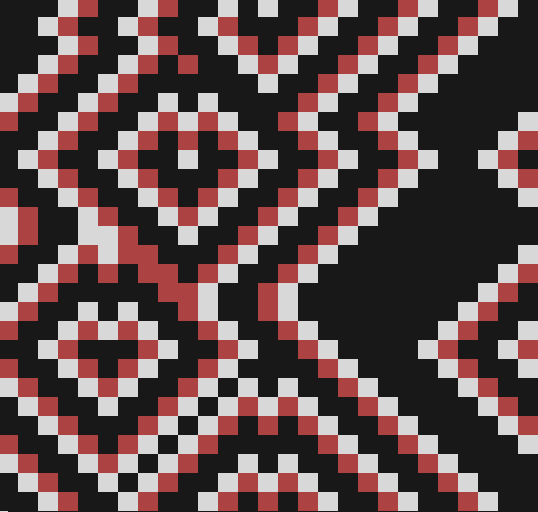}
    \includegraphics[width=0.24\linewidth]{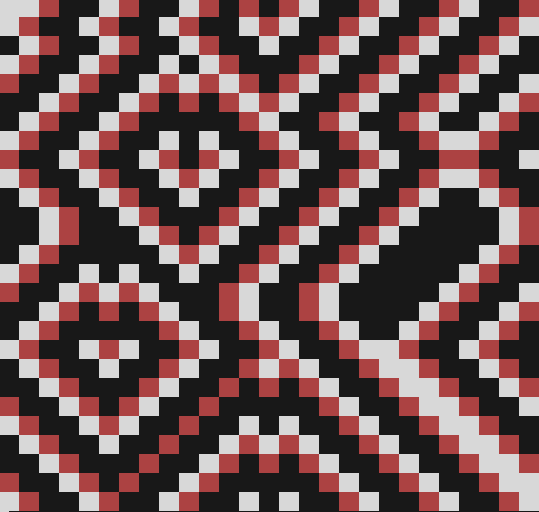} \includegraphics[width=0.24\linewidth]{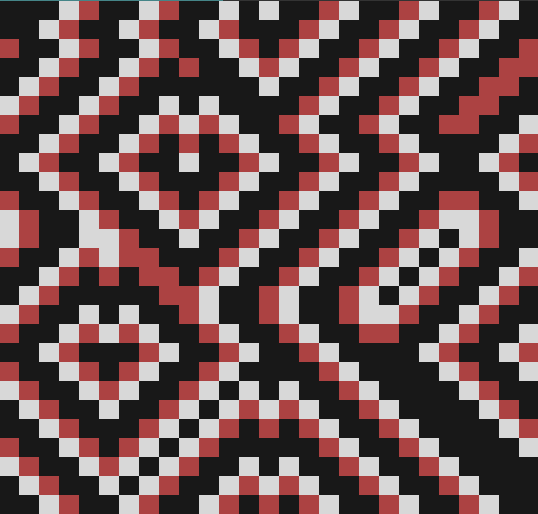} \includegraphics[width=0.24\linewidth]{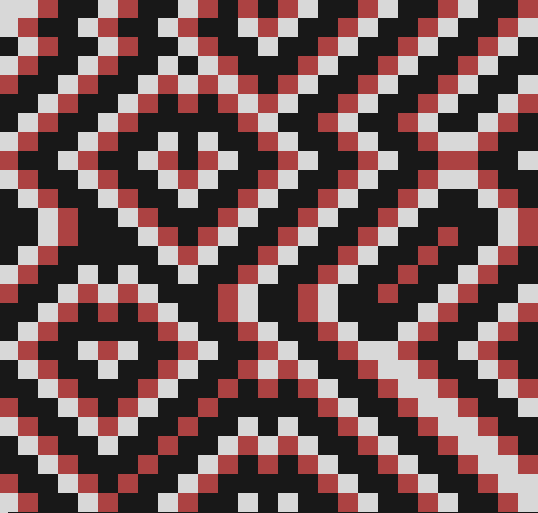} \includegraphics[width=0.24\linewidth]{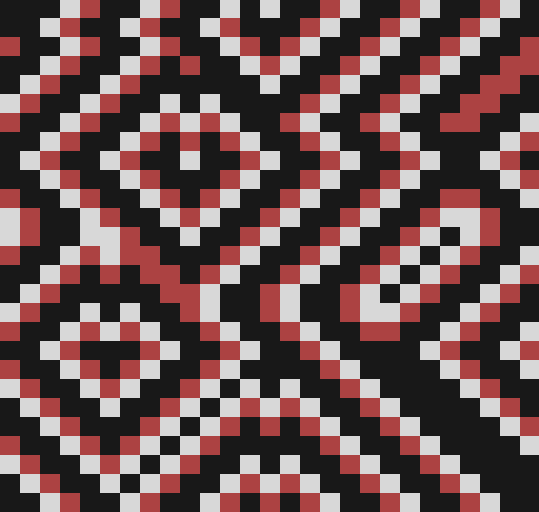}

    \caption{\textbf{Simulation of an excitable medium}.
Snapshots separated by  $\Delta t=2$ timesteps from a 27x27 system, using random initial conditions ($s=37873$) with probability of $0.1$ for cells to be spiking and $0.1$ for cells to be resting. After some irregular behaviour for 19 timesteps, the system settles into a periodic configuration of spirals, which in this case shows period 3.
    }
    \label{fig:excitable}
    \vspace{-1.5em}
\end{figure}

\subsection{Forest Fire}

\vspace{-0.5em}

The forest fire model is another famous CA which manifests self-organized criticality \citep{bak_chen_tang_1990}. The cells represent trees in a forest and, like excitable media, can take three states: \textsf{Fire}, \textsf{Tree}, and \textsf{Empty}. A tree ignites if there is a nearby burning tree and a burning cell becomes empty.

\begin{figure}[h!]
    \centering
    \includegraphics[width=0.24\linewidth]{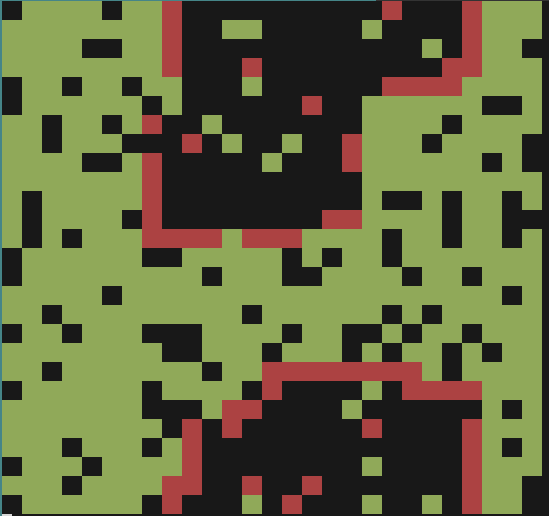} \includegraphics[width=0.24\linewidth]{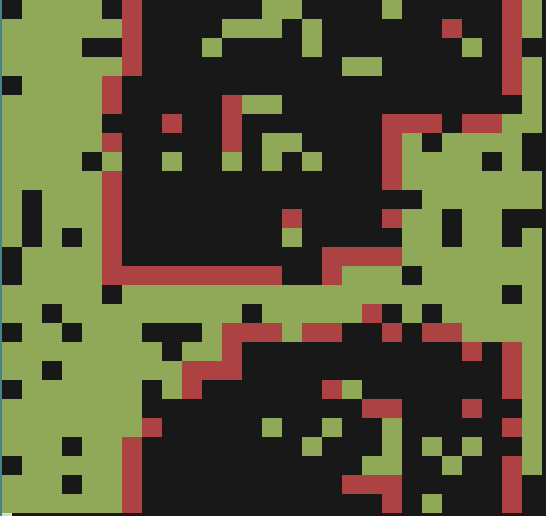} \includegraphics[width=0.24\linewidth]{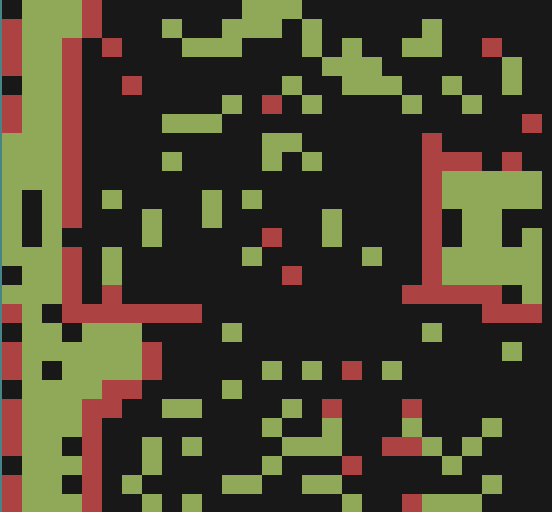} \includegraphics[width=0.24\linewidth]{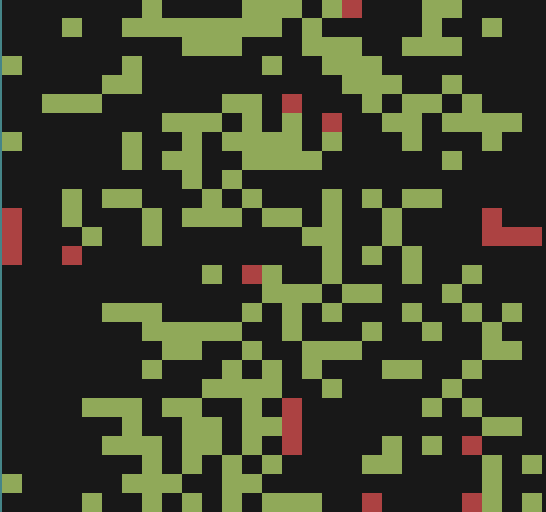}
    \caption{\textbf{Simulation of a stochastic Forest Fire model}.
Snapshots from a 27x27 system, separated by $\Delta t=2$ timesteps, evolving randomly (seed $s_1=4123$) with firing rate $f=0.0005$ and growing rate $p=0.1$, and random initial conditions ($s_1=1978$) given by the same rates. The balance between the growth rate and firing rate maintains the model in the critical state, where neither all trees have died nor do they survive forever.}
    \label{fig:forest}
\end{figure}

Unlike the CA presented so far, this model has a vital stochastic 
component: a tree can spontaneously ignite with a small probability $f$, while an empty cell can grow a new tree with probability $p$. The ratio between $f$ and $p$ controls system behaviour; at certain parameter ranges, the model enters a critical state, as seen in the example simulation above.

\section{From grids to networks}\label{sec:future}

Not only CA, but complex systems in general require local interactions between components to manifest emergence. As such, the comonadic model can apply to much more  complex simulations, as long as there is a distinct, fixed topology. One could show, by induction, that the comonadic array model in 1D and 2D can be easily expanded to arbitrary dimensions.

Moreover, it is tempting to extend the model to arbitrary topologies; while in the current implementation we only specified neighbourhoods as operations with indices on arrays, these could, alterntively, be expressed as neighbour lists on an underlying network. Such an implementation, alongside the support for environment variables and randomness, would allow modelling social networks or brain regions and emergent spreading phenomena like epidemics or information flow.

\section{Conclusion}\label{sec:concl}

We have presented an example comonadic implementation of cellular automata which, unlike existing implementations, uses an underlying array structure and monad and comonad transformers to add further environment variables and stochasticity in the comonadic evaluation.

This contribution is valuable as it offers an easily-parallelisable and extensible framework to simulate CA.
Thanks to the Curry-Howard-Lambek isomorphism, we may link this implementation to both a mathematical description of the model and its semantics.

This brings to mind Rine's categorical charaterisation of automata, offering a means to identify what an arbitrary system can do given a categorical description of its input and output states \citep{rine_1971}. In this way we can effectively provide a theory of the traits exhibited by a system. It will be interesting to consider the effect of this approach on theories of emergence and criteria for detecting emergence in general.
In conjuncture to recent descriptions of complex systems as computation structures (more specifically $\epsilon$-machines) \citep{rosas_2024}, it is extremely timely to turn to categorically-theoretical descriptions of multi-agent systems and explore the isomorphism between semantics, mathematics, and simulations, towards a unifying mathematical theory of complex behaviour emerging by computation. \\
\\

\section*{References}

\bibliographystyle{apalike}
\bibliography{bib}

@article{greenberg_hastings_1978, title={Spatial Patterns for Discrete 
Models of Diffusion in Excitable Media}, volume={34}, 
DOI={https://doi.org/10.1137/0134040}, number={3}, journal={SIAM J. Appl. Math.}, publisher={Society for Industrial and Applied 
Mathematics}, author={Greenberg, J M and Hastings, S P}, year={1978}, 
month={May}, pages={515–523} }

@article{wolfram_1983, title={Statistical mechanics of cellular automata}, volume={55}, DOI={https://doi.org/10.1103/revmodphys.55.601}, number={3}, journal={Rev. Mod. Phys.}, author={Wolfram, Stephen}, year={1983}, month={Jul}, pages={601–644} }

@article{bak_chen_tang_1990, title={A forest-fire model and some 
thoughts on turbulence}, volume={147}, 
url={https://www.sciencedirect.com/science/article/pii/037596019090451S},
 DOI={https://doi.org/10.1016/0375-9601(90)90451-s}, number={5-6}, 
journal={Phys. Lett. A}, author={Bak, Per and Chen, Kan and Tang, 
Chao}, year={1990}, month={Jul}, pages={297–300} }

@book{wolfram_2019, title={A New Kind of Science}, 
ISBN={9781579550257}, url={https://www.wolframscience.com/nks/}, 
publisher={Champaign, Il Wolfram Media}, author={Wolfram, Stephen}, 
year={2019}, pages={1087} }

@article{ciacci_falkenberg_2020, 
title={Understanding the transition from paroxysmal to persistent atrial
 fibrillation}, volume={2}, 
author={Ciacci, Alberto and others},
DOI={https://doi.org/10.1103/physrevresearch.2.023311}, number={2}, 
journal={Phys. Rev. Res.}, publisher={American Physical 
Society}, 
year={2020}, month={Jun} 
}

@article{rabinovitch_2024,
  author       = {Rabinovitch, A. and others},
  title        = {Ephaptic conduction in tonic–clonic seizures: A cellular automaton model},
  journal      = {Front. Neurol.},
  year         = {2024},
  volume       = {15:1477174},
  month        = {November},
  doi          = {10.3389/fneur.2024.1477174},
  url          = {https://www.frontiersin.org/articles/10.3389/fneur.2024.1477174/full}
}

@article{bedau_2008,
  title     = {Weak Emergence},
  author    = {Bedau, Mark A.},
  journal   = {Noûs},
  volume    = {31},
  pages     = {375-399},
  year      = {1997},
  publisher = {Blackwell Publishers}
}

@article{rosas_2024, 
title={Software in the natural world: A computational approach to emergence in complex multi-level systems}, 
doi={10.48550/arxiv.2402.09090},
author={Rosas, Fernando E and others},
volume={arXiv:2402.09090},
year={2024}, 
month={Feb} }

@book{jensen_2022,
	title        = {{Complexity Science}},
	isbn         = {9781108834766},
	publisher    = {Cambridge UP},
	author       = {Henrik Jeldtoft Jensen},
	year         = 2022,
	month        = {Oct}
}

@article{church1936,
  author       = {Alonzo Church},
  title        = {{An Unsolvable Problem of Elementary Number Theory}},
  journal      = {Am. J. Math.},
  volume       = {58},
  number       = {2},
  pages        = {345--363},
  year         = {1936},
  doi          = {10.2307/2371045},
  comment      = {Introduced the lambda calculus and formalized effective computability}
}

@article{turing1936,
  author       = {Alan M. Turing},
  title        = {{On Computable Numbers, with an Application to the Entscheidungsproblem}},
  volume       = {2},
  number       = {42},
  journal      = {Proc. London Math. Soc.},
  pages        = {230--265},
  year         = {1936},
  doi          = {10.1112/plms/s2-42.1.230},
  comment      = {Introduced the Turing machine and formalized computability}
}

@article{hoare1980,
  author       = {William P. Hoare},
  title        = {{The Correspondence Between Proofs and Programs}},
  journal      = {Commun. ACM},
  volume       = {23},
  number       = {7},
  pages        = {417--422},
  year         = {1980},
  doi          = {10.1145/358896.358927},
  comment      = {Early exposition of the Curry–Howard correspondence}
}

@incollection{Mitchell1990,
    title = {Type Systems for Programming Languages},
    editor = {Jan {van Leeuwen}},
    booktitle = {Formal Models and Semantics},
    pages = {365-458},
    year = {1990},
    series = {Handbook of Theoretical Computer Science},
    isbn = {978-0-444-88074-1},
    doi = {https://doi.org/10.1016/B978-0-444-88074-1.50013-5},
    url = {https://www.sciencedirect.com/science/article/pii/B9780444880741500135},
    author = {John C. Mitchell},
}

@article{rine_1971, title={A categorical characterization of general automata}, volume={19}, DOI={https://doi.org/10.1016/s0019-9958(71)80005-0}, number={1}, journal={Information and Control}, author={Rine, David C.}, year={1971}, month={Aug}, pages={30–40} }

@inproceedings{capobianco_uustalu_2010, title={A Categorical Outlook on Cellular Automata}, url={https://arxiv.org/abs/1012.1220}, 
booktitle={Proceedings of JAC 2010. Journées Automates Cellulaires},  author={Capobianco, Silvio 
and Uustalu, Tarmo}, year={2010}, pages={88–99} }

@article{uustalu_2008, title={Comonadic Notions of 
Computation}, volume={203}, 
DOI={https://doi.org/10.1016/j.entcs.2008.05.029}, number={5}, 
journal={Electron. Notes Theor. Comput. Sci.}, 
publisher={Elsevier BV}, author={Tarmo Uustalu and Vene, Varmo}, 
year={2008}, month={Jun}, pages={263–284} }

@article{widemann_hauhs_2011, 
title={Distributive-Law Semantics for Cellular Automata and Agent-Based 
Models}, ISBN={9783642229435}, 
DOI={https://doi.org/10.1007/978-3-642-22944-2_24}, journal={Lec. Notes Comp. Sci.}, publisher={Springer Berlin Heidelberg}, 
author={Widemann, Baltasar Trancón and Hauhs, Michael}, year={2011}, 
pages={344–358} }

@article{widemann_2012, title={Structural Operational Semantics for 
Cellular Automata}, DOI={https://doi.org/10.1007/978-3-642-33350-7_19}, 
journal={Lec. Notes Comp. Sci.}, publisher={Springer 
Science+Business Media}, author={Widemann, Baltasar Trancón }, 
year={2012}, month={Jan}, pages={184–193} }

@book{leinster_2017, title={Basic category theory}, ISBN={9781107044241}, url={https://arxiv.org/pdf/1612.09375}, publisher={Cambridge UP}, author={Leinster, Tom}, year={2017} }

@article{basold_2025, title={An Expressive Coalgebraic Modal Logic for Cellular Automata}, volume={arXiv:2504.16735}, author={Basold, Henning and Ford, Chase and Pirée, Lulof}, year={2025} }

@phdthesis{vanheerdt2020calf,
  author       = {Gerrit Kornelis van Heerdt},
  title        = {The Categorical Automata Learning Framework},
  school       = {University College London},
  year         = {2020},
  url          = {https://discovery.ucl.ac.uk/10110356/1/thesis_final_ucl.pdf},
}

@article{colcombet2020learning,
  author       = {Thomas Colcombet and Daniela Petrişan and Riccardo Stabile},
  title        = {Learning Automata and Transducers: A Categorical Approach},
  volume       = {arXiv:2010.13675},
  year         = {2020},
  url          = {https://arxiv.org/pdf/2010.13675.pdf}
}

@misc{kmett_2015, title={Cellular Automata}, howpublished={\href{https://www.schoolofhaskell.com/user/edwardk/cellular-automata}{schoolofhaskell.com}}, journal={School of Haskell}, publisher={Schoolofhaskell.com}, author={Kmett, Edward}, year={2015}, month={Jun} }

@misc{piponi_2006, title={ Evaluating cellular automata is comonadic }, howpublished={\href{http://blog.sigfpe.com/2006/12/evaluating-cellular-automata-is.html}{blog.sigfpe.com}}, journal={A neighborhood of Infinity}, publisher={Sigfpe.com}, author={Piponi, Dan}, year={2006}, month={Dec} }

@misc{potts_2006, title={{From Bits to Cells: Simple Cellular Automata in Haskell, Part Two}}, howpublished={\href{https://praisecurseandrecurse.blogspot.com/2006/12/from-bits-to-cells-simple-cellular_19.html}{praisecurseandrecurse.blogspot.com}}, journal={Praise, curse and recurse}, publisher={blogspot.com}, author={Potts, Paul R.}, year={2006}, month={Dec} }

@misc{bollu_2019_gh, title={{A collection of Cellular Automata written in Haskell with Diagrams}}, howpublished={\href{https://github.com/bollu/cellularAutomata}{github.com/bollu/cellularAutomata}}, journal={Github.com}, author={Bhat, Siddharth and Garay, Felipe }, year={2019} }

@misc{sas_sutherland_2019, title={Hascell}, howpublished={\href{https://github.com/mearlboro/hascell}{github.cm/mearlboro/hascell}}, author={Sas, Madalina I and Sutherland, Julian H J}, year={2024}}

@misc{kmett_2-14,
    author = {Edward Kmett},
    year = {2014},
    title = {{The Comonad Package}},
    howpublished={\href{hackage.haskell.org/package/comonad}{hackage.haskell.org}}
}

\end{document}